\documentclass[twocolumn]{revtex4-1}
\usepackage[T1]{fontenc}
\usepackage[latin1]{inputenc}
\usepackage{amssymb}
\usepackage{amsmath}
\usepackage{graphicx,wasysym}
\usepackage[pdftex]{hyperref}
\newcommand{\eq}[1]{Eq.~(\ref{#1})}

\newcommand{\N}{{\cal N}}
\pdfoutput=1

\makeatletter

\usepackage{epsf}

\def\be{\begin{equation}}
\def\ee{\end{equation}}
\def\ba{\begin{eqnarray}}
\def\ea{\end{eqnarray}}
\def\beas{\begin{eqnarray*}}
\def\eeas{\end{eqnarray*}}
\def\sla{\raise.15ex\hbox{$/$}\kern-.57em}

\begin{document}

\title{\Large\bf Measure Problem for Eternal and Non-Eternal Inflation}

\author{Andrei Linde}
\author{Mahdiyar Noorbala}
\affiliation{Department of Physics, Stanford University, Stanford, CA 94305, USA}


\begin{abstract}
We study various probability measures for eternal inflation by applying their regularization prescriptions to models where inflation is not eternal.  For simplicity we work with a toy model describing inflation that can interpolate between eternal and non-eternal inflation by continuous variation of a parameter.  We investigate whether the predictions of four different measures (proper time, scale factor cutoff, stationary and causal {diamond}) change continuously with the change of this parameter.  We will show that {only} for the stationary measure the predictions change continuously. For the proper-time and the scale factor cutoff, the predictions are strongly discontinuous. For the causal diamond measure, the predictions are continuous only if the stage of the slow-roll inflation is sufficiently long.
\end{abstract}

\maketitle

\section{Introduction}

Inflationary cosmology provides a simple mechanism which explains the observed homogeneity of our world: Inflation takes a tiny domain of the universe and rapidly expands it to the size which may exceed by many orders of magnitude the size of the observable part of the universe. This stretching removes all previously existing inhomogeneities and renders our world uniform. However, this mechanism does not make the universe {\it globally} uniform. If the universe from the very beginning consisted of different parts with different properties (e.g. the scalar fields occupying different minima of their potential energy), then the post-inflationary universe becomes divided into many exponentially large parts with different properties and even with different laws of the low-energy physics operating in each of them  \cite{linde1982,linde1983}. Moreover, even if initially the universe was represented by a single homogeneous domain, inflationary quantum fluctuations may divide it into many exponentially large parts with different properties. In effect, an inflationary universe becomes a multiverse consisting of exponentially large ``universes.'' This process leads to most profound consequences if inflation is eternal  \cite{Vilenkin:1983xq,Linde:1986fd,Linde:1987aa,Linde93,Linde94,GarciaBellido:1993wn,Vilenkin:1994ua}, but similar effects may occur even if inflation is not eternal \cite{Linde:1984je}.

These observations provided a simple scientific justification for the use of anthropic principle in inflationary cosmology. One should be much less surprised  to see that various parameters of the theory of elementary particles take non-generic, fine-tuned values if this is what makes our life possible. Our life is also non-generic, so if the universe provides us with the choice of generic vacua where we cannot live and non-generic ones where we can live, the choice is obvious. However, in order to use this argument to its full potential, and to go from ``possible'' to ``probable,'' one should learn how to compare the probabilities to live in different parts of the multiverse.

The main problem here is that in an eternally inflating universe the total volume occupied by all, even absolutely rare types of the ``universes,'' is indefinitely large. Therefore comparison of different types of vacua involves comparison of infinities. As emphasized already in the first papers on the probability measure in eternal inflation \cite{Linde93,Linde94,GarciaBellido:1993wn}, such a comparison is inherently ambiguous and depends on the choice of the cutoff, which is required to regularize the infinities. However, as we are going to see shortly, the measure problem may appear even if inflation is not eternal and the universe is finite. 

Historically, the first probability distribution considered in the literature was the function $P_{c}(\phi,t)$ \cite{Starobinsky:1986fx,Goncharov:1987ir}. It described the probability to find a given scalar field at a given time at a given point. One can equivalently interpret  $P_{c}(\phi,t)$  as the probability distribution in the comoving coordinates, which do not reflect the exponential expansion of the universe during inflation.

This probability distribution is inconvenient for the description of eternal inflation, which occurs because of the exponential growth of the parts of the universe remaining at the stage of inflation. Eternal inflation occurs even when it could seem improbable in terms of $P_{c}(\phi,t)$. As a result, an investigation of eternal inflation initially was performed with the help of a different probability distribution $P(\phi,t)$, which rewarded vicinity of each inflationary point by a factor reflecting the growth of its volume \cite{Linde:1986fd,Goncharov:1987ir}. An important advantage of this probability measure was the stationarity (time-independence) of the distribution $P(\phi,t)$ in the limit of large $t$ \cite{Linde93,Linde94}. This means, for example, that if one calculates the ratio of the volume of all parts of the universe containing the field $\phi_+$ to the volume of all parts with the field $\phi_-$  starting from some sufficiently large time $t$, then the result will not change when the time $t$ continues to grow.

However, the probability distribution $P(\phi,t)$ depends on the choice of time parameterization. If $t$ is the usual proper time, which is measured by the usual clock, then the vicinity of each point during inflation grows like $e^{3H \Delta t}$ during each small time interval $\Delta t$. This rewards expansion of the universe with large $H$. But one can also measure time in terms of expansion of space $\tau \sim a$, where $a$ is the scale factor \cite{Starobinsky:1986fx,Linde:1986fd,Goncharov:1987ir,Linde93,Linde94}, or, equivalently, in terms of $\eta \sim \log \tau = \log a$. In this case, vicinity of each point is uniformly rewarded by the $H$-independent factor $e^{3 \Delta \eta}$. More generally, one can introduce a family of measures where expansion is rewarded by a factor $e^{3H^{\beta} \Delta t_{\beta}}$, where $\beta$ is some constant depending on choice of time slicing $t_{\beta}$. Each of these choices is quite legitimate, and the results are stationary in the large time limit, but the results obtained by this method are exponentially sensitive to the choice of the cutoff, i.e. to the choice of $\beta$  \cite{Linde93,Linde94}.

Since that time, dozens of different candidates for the role of the probability measure have been proposed, most of them giving different predictions. It is impossible to give a full list of different proposals here, a partial list can be found  e.g. in \cite{Winitzki:2006rn}.  We will mention only few of them, which are often discussed now, and briefly discuss their advantages and problems.

Out of all of these measures, the original proper time cutoff measure $P(\phi,t)$ is the simplest. However, this measure suffers from the youngness problem \cite{Guth:2007ng}, which was especially clearly formulated in \cite{Tegmark:2004qd}: This measure exponentially rewards parts of the universe staying as long as possible at the highest values of energy density. As a result, this measure exponentially favors life appearing in the parts of the universe with an extremely large temperature, which contradicts the observational data.

The scale factor cutoff measure $P(\phi,\tau)$, which corresponds to the choice $\beta = 0$, does not suffer from the youngness problem because it does not give exponentially large rewards to the parts of the universe spending extra time at a large energy density. Therefore this measure, as well as its various modifications and generalizations, became quite popular lately, see e.g. \cite{DeSimone:2008bq,Bousso:2008hz,DeSimone:2008if,Garriga:2008ks,Bousso:2009dm}.

However, this advantage occurs because the scale factor cutoff measure corresponds to the special choice $\beta = 0$ in the class of measures where expansion is rewarded by $e^{3H^{\beta} \Delta t_{\beta}}$. For all values of $\beta >0$, these measures suffer from the youngness problem, and for $\beta < 0$ they suffer from the opposite problem, which can be called ``oldness'' problem: Life is predicted to exist mostly  in cold empty space. This is equivalent to the so-called Boltzmann brain problem. It appears in a certain class of measures predicting that typical observers should be created not as a result of the usual cosmological evolution, but because of quantum fluctuations in an empty post-inflationary universe \cite{Dyson:2002pf}. 

Thus the choice $\beta = 0$ must be made with an incredible precision. An optimist  may consider it as an indication that the special choice $\beta =0$ is preferable. A pessimist may counter it by saying that the scale factor cutoff measure provides an ultimate example of an exponential instability of predictions with respect to the choice of time parameterization, which seems unphysical.

Another possibility is to consider the causal diamond measure \cite{Bousso:2006ev,Bousso:2007kq} which, unlike the measures discussed so far, is not global and does not involve any time cutoff.  This measure cuts from the spacetime inside any given vacuum a finite four-volume subset (the causal diamond) formed by the intersection of the future light cone of the point where ``an observer'' crosses the hypersurface of reheating  and the past light cone of the point where he/she leaves that vacuum. This makes all regularized quantities finite within the causal diamond.  The measure then assigns (i) prior probabilities to vacua of each type based on the number of times when an observer appears in a given vacuum and (ii) weights to each vacuum according to the entropy produced in its causal diamond.  The net probability is given by the product of these two numbers.  The predictions obtained by this method depend on the initial probability distribution of the vacua. If, for example, one evaluates the probability of initial conditions using the Hartle-Hawking wave function, then the causal diamond measure suffers from a severe  Boltzmann brain problem \cite{Linde:2006nw}. Under many alternative assumptions concerning the probability distribution of initial conditions, the causal diamond measure gives practically the same results as the scale factor cutoff measure and the recently proposed measure based on considerations related to holography and conformal invariance \cite{Garriga:2008ks}.  When using the causal diamond measure in this paper we assume that the prior probabilities are uniform on different vacua.  We also take the number of galaxies in a causal diamond as a proxy for the amount of entropy produced there.

All of these measures share certain vulnerability with respect to the Boltzmann brain problem. Roughly speaking, in order to avoid this problem for these measures, the decay rate of each of the vacua in the landscape  must be greater than the rate of the Boltzmann brain production there. For a more exact formulation of this  condition see \cite{Bousso:2008hz,DeSimone:2008if}. This condition may be satisfied for a rather broad class of vacua in string theory landscape \cite{Freivogel:2008wm}, but the answer for generic string theory vacua is not known yet. And here lies a potential problem: If there is a 50\% probability that the required condition is satisfied  in {\it each} of the exponentially large number of the as yet unexplored stringy vacua (we made the 50\% probability assumption simply because we do not really know the answer), then the probability that it is satisfied in {\it all} of the unexplored vacua is exponentially small.

Potential difficulties of this class of measures are quite considerable, but they are less severe than the youngness problem  plaguing the proper time cutoff  measure $P(\phi,t)$. Fortunately, it was possible to address the youngness problem in an improved version of this measure, which was called the stationary measure \cite{Linde:2007nm,Linde:2008xf}. The main observation of Ref.~\cite{Linde:2007nm} was that the stationary regime, in which $P(\phi,t)$ becomes time-independent, is established at different times for different processes. This means that there there was no stationarity in the beginning of the process. Therefore it does not make sense to compare all processes at the same time. Instead of that one should compare different processes starting at the time when the stationarity is first achieved for each of them separately.

For example, in  the model to be studied in our paper (see Figs. 1, 2), we will try to compare the volume of all parts of the universe with the scalar field $\phi_+$ to the volume of all parts with the  field $\phi_-$. In the beginning of the process the universe was in a state $\phi = 0$. Then, after the tunneling and slow roll, the domains of the field $\phi_-$ were produced, and their number began growing exponentially starting from some time $t_{i-}$. The domains of the field $\phi_+$ were produced only somewhat later, and their number began growing at the same rate starting from a different time $t_{i+}$. The main idea of the stationarity measure  \cite{Linde:2007nm,Linde:2008xf} is that one should compare the domains with the field  $\phi_+$ to the domains with the field $\phi_-$ not at the same time $t$ after the big bang, but at the same time $\Delta t$ since the moment when each of the domains were produced and their number started growing at the same rate.

This idea has a particularly simple interpretation  in the situations when inflation is not eternal. Indeed,  in this context it does not make any sense to compare processes in different domains at the same time from the big bang. For example, if one uses the scale factor measure, the process of star formation begins at exponentially different times (i.e. at different scale factors)  in different parts of the universe. However, it is quite possible to compare processes of star formation in different domains starting from the time when the star formation actually begins there. This is the main principle of the stationarity measure:  synchronization in terms of an arbitrary choice of time should be replaced by synchronization with respect to some kind of a physical process  \cite{Linde:2007nm,Linde:2008xf}.

One of the advantages of this measure is the absence of the exponential sensitivity of predictions to the choice of the time parameterization \cite{Linde:2007nm}. An investigation performed in \cite{Linde:2008xf} suggests that this measure does not suffer from the youngness problem and  the Boltzmann brain problem.  

Stationary measure does not reward us for growth of volume during a purely de Sitter stage in a metastable vacuum state (i.e. in false vacuum). However, the probabilities are proportional to the growth of volume during the stage of the slow roll inflation. This property explains flatness of the universe. There is a potential danger that the exponential sensitivity of the total volume of the universe to the choice of the inflationary parameters may make the total number of observers in the universe exponentially large, but simultaneously make the probability of emergence of life in any finite volume extremely small. This is the essence of the problem which is sometimes called Q catastrophe \cite{Vilenkin:1994ua,Feldstein:2005bm,Garriga:2005ee}. Possible solutions of this problem were proposed in 
\cite{GarciaBellido:1994ci,Linde:2005yw,Hall:2006ff}.

This brief discussion illustrates the general situation with the probability measure for eternal inflation.  In this paper we will discuss the situations when inflation is not eternal and the universe is compact. As we will see, even in such situations one may come to different conclusions with respect to probabilities. Therefore it may make sense to temporarily suppress our ambitions and try to understand non-eternal inflation, which at the first glance could seem quite trivial, and then return again to the investigation of eternal inflation. We will see that the stationary probability measures lead to similar predictions for eternal and non-eternal inflation. However, all other measures discussed above give very different predictions for models with eternal and non-eternal inflation. While this discontinuity does not necessarily mean that such measures are problematic, we think that this fact requires certain attention.

\section{Eternal and Non-eternal Inflation: A Toy Model}

\begin{figure}[t]
\centering
\includegraphics[scale=1.7]{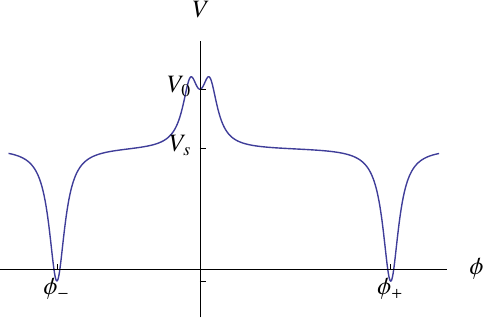}
\caption{The scalar field potential for the eternal inflation model.}
\label{fig:EtPot}
\end{figure}

We consider inflation driven by a scalar field in models with two different potentials.  In one model (which we call the eternal model; see Fig. \ref{fig:EtPot}), the scalar field starts at $\phi_0$. Inflation proceeds via tunneling through the potential barriers to the right or left with an equal rate $\Gamma$ per proper volume.  Bubbles of new vacua nucleate and undergo a subsequent slow roll inflation along the fairly flat shoulders of the potential in either side.  The value of the potential $V_s$ on these shoulders is the same but the field excursion along them is different.  Eventually inflation ends and the scalar field takes on the value $\phi_+$ or $\phi_-$ in the corresponding part of the universe.  We assume, for simplicity, that these are vacua with the same particle physics and the same vacuum energy densities $\Lambda<0$, which collapse in a finite time.  As long as the expansion rate $3H_0$ of the volume populated by false vacuum is greater than its total decay rate $\frac{8\pi}{3}H_0^{-3}\Gamma$ (which is true in most realistic situations), inflation is eternal.  The space-time takes a fractal structure with infinite 4-volume and a measure is required to make any statistical statement.  

\begin{figure}[t]
\centering
\includegraphics[scale=1.7]{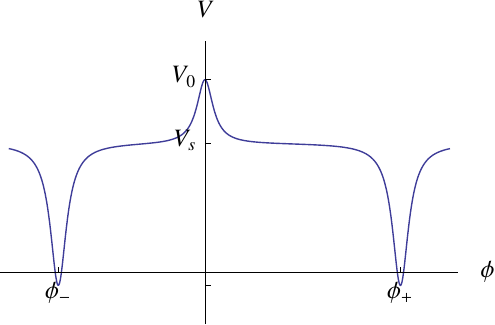}
\caption{The scalar field potential for the non-eternal inflation model.}
\label{fig:NonEtPot}
\end{figure}

In the other model (the non-eternal model; see Fig. \ref{fig:NonEtPot}) the scalar field starts from $\phi_0$ again but there is no potential barrier and, instead of tunneling, the field just falls down to either right or left (again with equal probabilities since the shape of the potential is symmetric).  This part of the potential is assumed to be steep enough that quantum fluctuations cannot cause an upward jump and hence inflation is non-eternal.  Finally there is a subsequent slow roll inflation, like in the eternal model, ending in one of the collapsing vacua.  

In both cases, we will make a simplifying assumption that the different parts of the universe experience a long stage of inflation, so that each such part becomes exponentially large,  and all interesting processes in each of these parts occur practically independently, as if they were separate universes. Of course, this picture is only approximately correct, and one should be careful not to use it beyond the limits of its applicability  (we will discuss the situation where this assumption breaks in Section~\ref{sec:non-et}). We hope that this approximate picture will be sufficient to identify some important differences which appear when one tries to describe models of eternal and non-eternal inflation.

Even if inflation is not eternal, one may still face the problem of regularizing infinities if, for example, we discuss an open or flat spatially infinite universe. There are two ways to avoid it. The most obvious way is to consider a closed universe with $\Lambda \leq 0$. It has finite size and it collapses in finite time. Another, less trivial possibility is to consider a compact open or flat universe, the simplest example being a flat universe with periodic boundary conditions, represented by a compact $k=0$ 3-torus. Whereas this possibility may seem a bit unusual, an investigation of the probability of quantum creation of such universes shows that their formation is exponentially more probable than formation of closed universes \cite{Zeldovich:1984vk,Coule:1999wg,Linde:2004nz}. Therefore in this paper we will concentrate on the models describing a flat or open compact universe. However, all results will remain valid for the close universe case if the duration of inflation in a close universe is long enough to make it effectively flat.

The evolution of the scale factor after the field falls down the hill but before it begins the slow roll stage is the same on the right and left shoulders of the potential since it is symmetric in the vicinity of $\phi_0$.  This stage occurs in a similar way  for eternal and non-eternal inflation.  In the eternal model the field tunnels from the false vacuum $V_0$ to an escape point close to the top of the potential, and then it starts falling down the hill with zero initial velocity.  In the non-eternal model the field just falls down from the same height $V_0$ in the potential of a very similar shape, with zero initial velocity.  In both cases it takes the same time for the field to reach the terminal velocity $\dot{\phi}=-V'_s/3H_s$, where $3H_s^2=V_s$ is the vacuum energy density during the slow roll inflation and $V'_s$ is the slope of the potential during this stage. For simplicity we assume that $V_0$ and $V_s$ have the same order of magnitude, the scalar potential does not change much during inflation, and the slope of the potential is very small and nearly constant. 

The field then rolls for a time
\begin{equation}\label{tipm}
t_{i\pm} = \left| \frac{\sqrt{3V_s}(\phi_{\pm} - \phi_0)}{V'_s} \right|
\end{equation}
until it reaches the right or left minimum, respectively.  The stage of slow roll inflation and the subsequent reheating is the same for the two models.  The only difference is that in the eternal model the spatial geometry of the bubble interior is a $k=-1$ hyperboloid (open inflation) while in the non-eternal model the newly formed domain can be either open or flat \cite{Zeldovich:1984vk,Coule:1999wg,Linde:2004nz}. This difference is rather inessential since soon after the beginning of inflation the universe becomes flat. Thus, for the purposes of our paper one can assume that the universe from the very beginning was expanding exponentially, $a(t) \sim a_0\, e^{H_s t}$, and the total number of e-folds of inflation is given by $N_\pm = H_st_{i\pm}$, {where $0<t<t_{i\pm}$ is the time duration of inflation on right and left, respectively.}

The total size of the universe at the end of inflation $a(t_{i\pm}) \sim a_0\, e^{N_{\pm}}$ depends on the size of the universe $a_0$ at the beginning of inflation. We will assume that the size of the universe at the beginning of inflation was very small. It can be as small as the Planck length,  $a_0 \sim 1$, or it may be of the order of the inverse Hubble constant during inflation, $a_0 \sim H_s^{-1}$, or it may take an intermediate value $a_0 \sim H_s^{-1/3}$, starting from which the classical description of a compact flat universe becomes possible \cite{Linde:2004nz}. However, even a very large difference between various possible values of $a_0$ can be compensated by a slight change of $N_\pm$. For the purposes of our paper we will simply take $a_0 = 1$, and treat $N_\pm$ as input parameters (instead of the details of the potential).

At the end of the slow roll inflation, reheating takes place, which we assume to be instantaneous.  It produces radiation and matter with respective densities $\rho_{ri}$ and $\rho_{mi}$ in a universe with a small negative cosmological constant, $|\Lambda| \ll V_s$.  Energy conservation at $t=t_i$ requires that
\begin{equation}
V_s = \rho_{mi} + \rho_{ri} + \Lambda.
\end{equation}
The ratio $\rho_{mi}/\rho_{ri}$ of produced matter to produced radiation is determined by the particle physics which is the same on right and left.  Since $V_s$ and $\Lambda$ are fixed throughout this paper, we conclude that $\rho_{mi}$ and $\rho_{ri}$ are also constant parameters independent of model (i.e., of eternal or non-eternal nature of the model) and of $\pm$ (i.e., of right or left minimum of the potential).  Such a universe will reach a maximum size at a turning time $t=t_t$, then begins contracting and finally collapses to a singularity at the finite time $t=t_c$.  To see this we note from the FRW equation
\begin{equation} \label{FRWeq}
H^2 + \frac{k}{a^2} = \frac{1}{3} \left[ \rho_{mi} \left( \frac{a_i}{a}
 \right)^3 + \rho_{ri} \left( \frac{a_i}{a} \right)^4 + \Lambda \right],
\end{equation}
that for $\Lambda<0$ and $\rho_m, \rho_r \geq 0$ there is always a turning point $H_t=0$ after which a phase of contraction starts leading to a singularity at $a=0$.  

\begin{figure}[t]
\centering
\includegraphics[scale=.33]{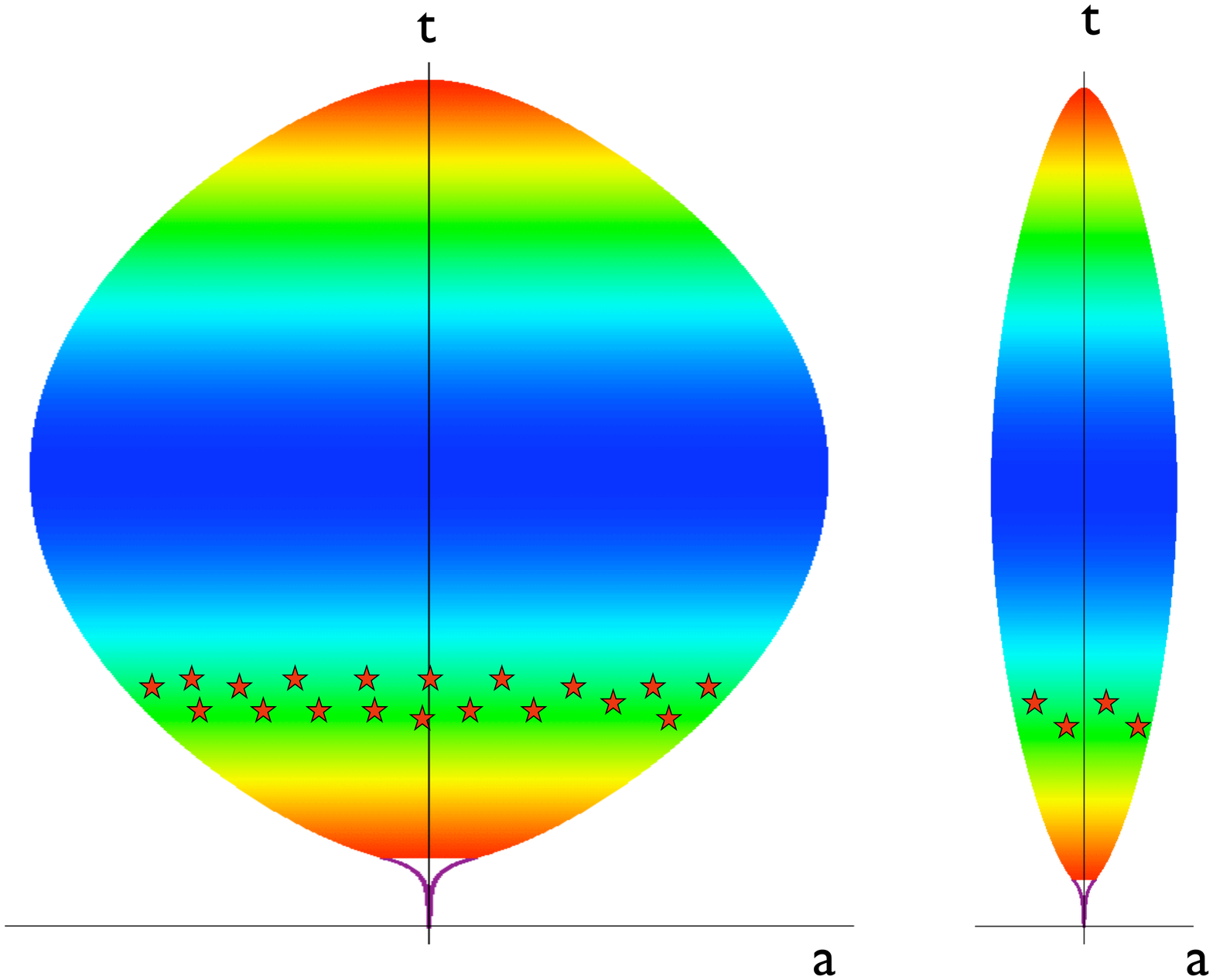}
\includegraphics[scale=.33]{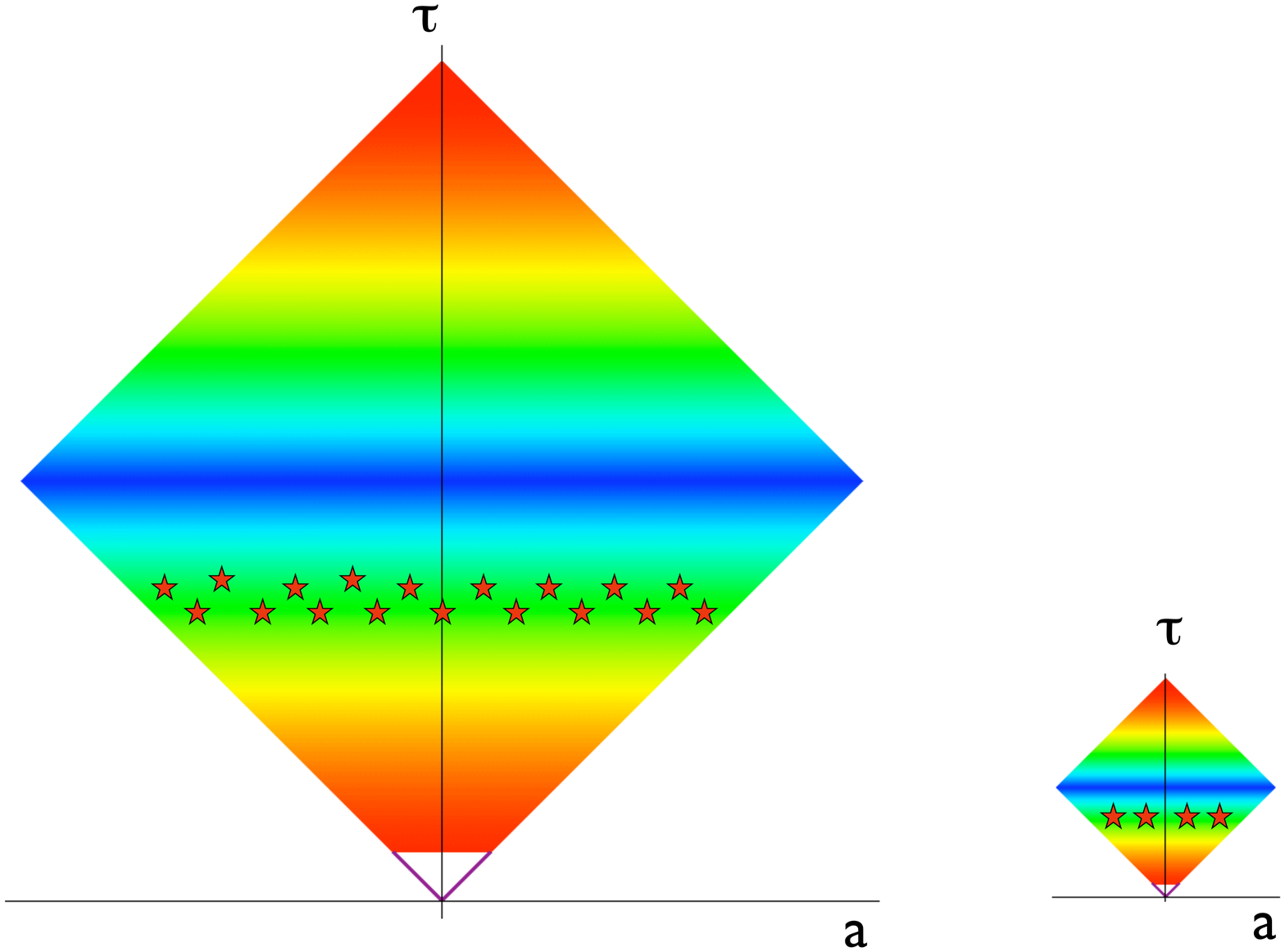}
\caption{Evolution of the scale factor in a collapsing $k=0$ FRW universe with respect to the proper time (top) and the scale factor time (bottom).  In each pair of plots the left one has a slightly longer stage of slow role inflation than the right one.  The area after reheating is painted. Notice almost exact symmetry of behavior of the scale factor during expansion versus contraction. Stars indicate the period where galaxies are produced and life as we know it is possible. This happens at the same time $t$ but at very different times $\tau$ for the universes experiencing longer/shorter stages of inflation.}
\label{fig:a}
\end{figure}

To estimate the total lifetime of the universe, let us first assume that inflation was long enough to render the term $ \frac{k}{a^2}$ irrelevant. (This term does not appear for the flat compact universe anyway.) Ignoring the duration of the period of inflation  ($t_i \ll t_c$), one can show that the total lifetime of the universe from its creation to its collapse is given by
\be
t_c  = 2 \int_0^{1}  \frac{\sqrt{3} {\alpha} d\alpha}{\sqrt{\Lambda \alpha^4 + \rho_{mt}\alpha + \rho_{rt}}} \label{tc} \ .
\ee
Here $\alpha$ is the ratio of the scale factor to its value at the turning point, and $ \rho_{mt}$ and $\rho_{rt}$ are the densities of matter and radiation at the turning point, where $ \rho_{mt}+\rho_{rt} + \Lambda = 0$.
 It is instructive to consider two separate cases: hot universe, $ \rho_{mt} = 0$, $\rho_{rt} + \Lambda = 0$, and cold universe, $\rho_{rt} = 0$, $\rho_{mt} + \Lambda = 0$. {Analytic} integration of this equation shows that for the hot universe
 \be
t_c^{\rm hot}  = {\frac{\pi}{2} \sqrt{\frac{3}{|\Lambda|}}}  \label{tc1} \ ,
\ee
and for the cold universe 
 \be
t_c^{\rm cold}  = {\frac{2\pi}{3} \sqrt{\frac{3}{|\Lambda|}}}  \label{tc2} \ .
\ee

Now let us restore the term $\frac{k}{a^2}$ in our equations and consider an open universe, ignoring $ \rho_{mt}$ and $\rho_{rt}$. In this case, 
 \be
t_c^{\rm open}  = {\pi \sqrt{\frac{3}{|\Lambda|}}}  \label{tc3} \ .
\ee
{More generally, if $k=0$ the collapse time is bounded from below by $t_c^{\rm hot}$ and from above by $t_c^{\rm cold}$.  If $k=-1$ the upper bound is $t_c^{\rm open}$ and the lower bound is $t_c^{\rm hot}$.}  In all of these cases the lifetime of the universe does not depend on the duration of inflation and is given by  $c |\Lambda|^{-1/2}$, where the coefficient $ c = O(1)$ depends on the matter contents of the universe. The only exception from this simple rule appears if one considers a closed universe with a short stage of inflation and an exponentially small cosmological constant. Then the universe collapses at a time shorter than $O(|\Lambda|^{-1/2})$. As we already mentioned, in this paper we will concentrate, for simplicity, on open or flat compact universes, or on closed universes with a sufficiently long stage of inflation, when the rule described above holds. However, most of the qualitative conclusions of our paper will remain valid for closed universes with a short inflationary stage.

We are interested in counting the total number $\N_{\pm}$ of galaxies across the whole spacetime, which are located inside a $\phi_{\pm}$ vacuum.  In a part of the universe where inflation has ended, galaxies are formed at a time $ t_g \gg t_i$ when the scale factor is $a_g=a(t_g)$.  As long as $\Omega \approx 1$, $ t_g$ must be independent of the inflationary history of universe and in particular of the number of $e$-folds (as long as the galaxies are much smaller than the total size of the universe).  The time $t_g$, and, consequently, the ratio $a_g/a_i$, and the number density of galaxies $n(t_g)$ can depend only on the particle physics of the vacuum, so they are $\pm$-independent.  

\section{The Eternal Inflation Model}

In this section we calculate the relative probability $\frac{\N_+}{\N_-}$ in the eternal model.  A careful analysis of the problem within the three cutoff-based measures requires solving a master equation which incorporates both the tunneling process and the diffusion equation that describes the stochastic behavior of the field during slow roll inflation \cite{Garriga:1997ef,Linde:2006nw,Linde:2007nm}.  We will not get into the details of this analysis, rather take a shortcut to the answer and explain our simplifying assumptions.

Let us consider a time cutoff $\tau$ that is related to the proper time $t$ via $d\tau = H^{1-\beta}dt$.  The total volume $V_0$ in the false vacuum state $\phi_0$ grows because of the exponential expansion of de Sitter space, but also slightly decreases due to the false vacuum decay. Ignoring the probability of jumps back to the original state $\phi_{0}$, one can write an equation for $V_0$:
\begin{equation}
\frac{dV_0(\tau)}{d\tau} = (3H_0^\beta - 2\kappa) V_0(\tau) \ .
\end{equation}
Here $\kappa = \frac{4\pi}{3} H_0^{\beta-4} \Gamma$ is the decay rate per unit $\tau$.  This equation has a simple exponential solution:
\begin{equation}
V_0(\tau) = V_0(0) \exp [(3H_0^\beta - 2\kappa) \tau].
\end{equation}

The second half of the process is the slow roll from the escape point(s) of tunneling to $\phi_\pm$.  We ignore the diffusion due to the quantum fluctuations and only consider the classical roll.  Then the volume $V_\pm$ of the part of the universe that has reached the time of galaxy formation on right/left is determined by
\begin{equation} \label{dVpm/dt}
\frac{dV_\pm(\tau)}{d\tau} = \kappa e^{3N_\pm} \left( \frac{a_{g\pm}}{a_{i\pm}} \right)^3 V_0(\tau - \tau_{g\pm}),
\end{equation}
where $\tau_{g\pm}$ is the time from bubble nucleation to galaxy formation:
\begin{equation} \label{taug}
\tau_g = \int d\tau = \int \frac{da}{H^\beta a}.
\end{equation}
In the special case of proper time one finds $\tau_g = t_g$, whereas for the scale factor time one gets $\tau_g = N + \log \frac{a_g}{a_i}$.

Noting that $a_g/a_i$ is $\pm$-independent and $N_\pm = H_s t_{i\pm}$ we find for the proper time ($\beta=1$)
\begin{equation}
\lim_{\tau\to\infty} \left. \frac{V_+(\tau)}{V_-(\tau)} \right|_{\beta=1} = \exp \left[ \left( 3 - \frac{3H_0 - 2\kappa}{H_s} \right) (N_+ - N_-) \right].
\end{equation}
The term $3H_0/H_s$ dominates the first factor in the exponent and leads to favoring of shorter inflation: the probability to be in $\phi_+$ is exponentially suppressed \cite{Linde:2007nm}.  For the scale factor time ($\beta=0$), however, we find:
\begin{equation}
\lim_{\tau\to\infty} \left. \frac{V_+(\tau)}{V_-(\tau)} \right|_{\beta=0} = e^{2\kappa (N_+ - N_-)},
\end{equation}
indicating a mild favoring of longer inflation (which is actually the source of the mild Boltzmann brain problem in this measure \cite{DeSimone:2008if}).  But since $\kappa$ is exponentially small, for the purposes of our comparison we can just consider the limit $\kappa \to 0$ which gives equal likelihood to $\phi_\pm$,
\begin{equation}\label{scale}
\lim_{\tau\to\infty} \left. \frac{V_+(\tau)}{V_-(\tau)} \right|_{\beta=0} = 1
\end{equation}

  To find the result for arbitrary $\beta$ one can break the integral in \eq{taug} into two pieces: $a<a_i$ and $a>a_i$.  The post inflation piece, $a>a_i$, is $\pm$-independent (similar to the integrals appearing in Eqs.~(\ref{tc}) and (\ref{CDo})) so it cancels out in the ratio while the inflationary piece, $a<a_i$ gives a contribution $\approx (N_+ - N_-)/H_s^\beta$.  Therefore one finds
\begin{equation}
\lim_{\tau\to\infty} \frac{V_+(\tau)}{V_-(\tau)} = \exp \left[ \left( 3 - \frac{3H_0^\beta - 2\kappa}{H_s^\beta} \right) (N_+ - N_-) \right].
\end{equation}

In the stationary measure one synchronizes the exponential growth of the volume of the two sides based on the time they reach the stationarity regime.  This amounts to modifying \eq{dVpm/dt} to read:
\begin{equation}
\frac{dV_\pm(\tau)}{d\tau} = \kappa e^{3N_\pm} \left( \frac{a_{g\pm}}{a_{i\pm}} \right)^3 V_0(\tau),
\end{equation}
whose solution yields:
\begin{equation}\label{stat}
\lim_{\tau\to\infty} \left. \frac{V_+(\tau)}{V_-(\tau)} \right|_{\hbox{\tiny stationary}} = e^{3(N_+ - N_-)}.
\end{equation}
For a more detailed derivation of this result see \cite{Linde:2007nm}.

All results obtained above describe the ratio $\frac{V_+(\tau)}{V_-(\tau)}$ at a given time $\tau$, if this time is large enough. The same results describe the ratio of all galaxies which ever existed in the universe until the cut-off time $\tau$, in the limit $\tau \to \infty$. This observation will play an important role when we will discuss possible generalizations of these results for the non-eternal inflation.

Finally for the causal diamond measure one finds the same result as in (\ref{scale}): \ $\lim_{\tau\to\infty} \left. \frac{V_+(\tau)}{V_-(\tau)} \right|_{\beta=0} = 1$, i.e. the final result does not depend on the duration of the slow roll inflation. Indeed, in our model the total lifetime of the universe, and, consequently, the total size of the causal diamond and the entropy produced there, do not depend on the duration of inflation. In the next section we will explain this result in a more detailed way and show that it remains valid for the non-eternal inflation as well, but only if inflation is long enough.

\section{The Non-eternal Inflation Model} \label{sec:non-et}

In the non-eternal model the total proper volume of all collapsing universes at their time of galaxy formation is finite, and so is the total number of galaxies.  This makes the calculation of the total number of galaxies simple and unambiguous. However, one immediately realizes some additional problems which were swept under the carpet in the eternal inflation scenario.

Indeed, in the eternal inflation scenario, galaxies of a given type will always exist in some parts of the universe at any time $\tau$, and the ratio of their number in different vacua asymptotically becomes time-independent. This is not the case for the non-eternal inflation scenario.
One can see it most clearly in the lower panel of Fig. 3, which shows two ``diamonds'' describing the evolution of the universe with respect to the scale factor time.  Whereas the proper time required for the galaxy formation in these universes is approximately the same, the corresponding scale factor time is dramatically different. As a result, it simply does not make any sense to compare the number of galaxies at the same  scale factor time $\tau$. 

This problem does not appear if one uses an analogue of the stationary measure, where the comparison occurs not at a given cosmological time, but at the time when certain physical processes (i.e galaxy formation) take place. The corresponding result in the stationary measure is proportional to the total volume of the universe at the time of the galaxy existence, which, in its turn, is proportional to the growth of volume during the slow roll inflation,
\begin{equation}\label{cutoff-NE}
\frac{\N_+}{\N_-} = \frac{a_{i+}^3}{a_{i-}^3} = \exp[3(N_+ - N_-)].
\end{equation}
This result coincides with the result which we obtained using the stationary measure in the eternal inflation case, see Eq.~(\ref{stat}).

An alternative approach is to compare the total number of galaxies which may be formed over the whole lifetime of the universe. This method works not only for the stationary measure, but for the proper time measure and for the scale factor measure. On average, $\phi_\pm$ domains are formed at the same time and each occupies half of the volume of the original false vacuum.  Thus we can find the ratio $\frac{\N_+}{\N_-}$ of all galaxies in $\phi_+$ to all of those in $\phi_-$ by simply computing the ratio $\frac{\N^{(1)}_+}{\N^{(1)}_-}$ of the galaxies in only one $\phi_+$ domain to those in one $\phi_-$ domain.  Ignoring geometric factors such as $4\pi/3$, one can write   $\N^{(1)}_{\pm} = n_{\pm}(t_{g\pm})\ a_{g\pm}^3$. As we saw earlier, $n_g$ and $a_g/a_i$ are $\pm$-independent.  Therefore, the ratio of the total number of galaxies which ever existed in the $\phi_+$ vacuum to those in the $\phi_-$ vacuum is again given by Eq.~(\ref{cutoff-NE}).
Once again, we recovered the answer which was obtained using the stationary measure for the eternal inflation scenario.

As already mentioned, the spacetime 4-volume is finite in this model and there is no need for a cutoff to regularize infinities.  Thus \eq{cutoff-NE} is an unambiguous result of counting of all galaxies in the universe. However, if one decides to impose an additional cutoff, which is not necessary for regularization of infinities in a compact universe, one may come to a different conclusion.  

In particular, the causal diamond measure does not take into account the parts of the universe that do not belong to the causal diamond.  Let us find the ratio $\N_+/\N_-$ given by this measure.  Consider an observer whose worldline crosses the reheating surface and hits the singularity far away from the walls, so that an FRW description with the observer at the center of coordinates is valid.  We need to count galaxies formed at $t_g$ but only those which lie inside the causal diamond.  The diamond has two boundaries: the past light cone of the point where the worldline hits the singularity $r = -\int_{t_c}^t \frac{dt}{a(t)}$, and the future light cone of the point where it hits the reheating surface $r = \int_{t_i}^t \frac{dt}{a(t)}$.  At $t_g < t_t$ the smaller of these two is the latter.  Thus at the time of galaxy formation $t_g$ there are
\ba 
\N_{\rm CD} &=& n_g \left( a_g \int_{t_i}^{t_g} \frac{dt}{a(t)} \right)^3 \nonumber \\
&=& n_g \left( \frac{a_g}{a_i} \int_1^{\frac{a_g}{a_i}} \frac{\sqrt{3} d\alpha}{\sqrt{\Lambda \alpha^4 + \rho_{mi}\alpha + \rho_{ri}}} \right)^3 \label{CDo}
\ea
galaxies within the causal diamond.  All variables appearing in this expression are $\pm$-independent and thus so is $\N_{\rm CD}$.  Furthermore, a uniform initial distribution of worldlines implies that on average an equal number of worldlines enter the $\phi_\pm$ domains and since they are far away from the walls they will stay in their domain until they fall into the singularity.  Thus $\N_\pm \propto \N_{\rm CD}$ with a $\pm$-independent proportionality constant.  Therefore, for the causal diamond measure we find
\begin{equation} \label{CD-NE}
\frac{\N_+}{\N_-} = 1.
\end{equation}
Thus the causal diamond measure, unlike all other measures discussed above, does not reward us for the exponential growth of volume during a long stage of slow roll inflation.  This result is valid for eternal and non-eternal inflation, but only if the stage of inflation is very long.

\begin{table*}[ht!]
{\small
\hfill{}
\begin{tabular}{|l|l|c|c|c|c|c|c|c|}
\hline
{\bf ~~~~~~~~~~~~ Measure  ~~~~~~~~~~~~} & {\bf ~~~~~~~~~~~~ Non-Eternal Model ~~~~~~~~~~~~} & { \bf ~~~~~~~~~~~~ Eternal Model ~~~~~~~~~~~~} \\ \hline \hline
Proper Time & ~~~~~~~~~~~~$\exp[3(N_+ - N_-)]$ & $\exp \left[ -\frac{3H_0}{H_s} (N_+ - N_-) \right]$ \\ \hline
Scale Factor &~~~~~~~~~~~~ $\exp[3(N_+ - N_-)]$ & 1 \\ \hline
Causal Diamond & $\begin{cases} \exp[3(N_+ - N_-)] & \hbox{if $N$ is small enough}\\ ~~~~~~~~~~~~1  & \hbox{if $N$ is large enough} \end{cases}$ & 1 \\\hline
Stationary &~~~~~~~~~~~~ $\exp[3(N_+ - N_-)]$ & $\exp[3(N_+ - N_-)]$ \\ \hline
\end{tabular}}
\hfill{}
\caption{Predictions of the four measures in the eternal and non-eternal models.}
\label{tab:1}
\end{table*}

Indeed, if inflation is very long, any observer will be able to see (and to influence) only a tiny fraction of the universe determined by the size of the causal diamond, which does not depend on the duration of inflation. However, if inflation is short and the universe is small enough,  an observer will be able to see all galaxies in the universe. In this case the total number of galaxies accessible to observations will be proportional to the total volume of the universe, and instead of Eq.~(\ref{CD-NE}) one should use Eq.~(\ref{cutoff-NE}), which shows that the probability depends exponentially on the duration of inflation, if it is short enough.

\section{Conclusions}

We have investigated four different measures in an eternal and a non-eternal toy model.  Table~\ref{tab:1} summarizes  the predictions of each measure for the relative probability of being in $\phi_+$ versus being in $\phi_-$.  In terms of rewarding for a long stage of inflation we have found that:
\begin{itemize}
	\item The proper time (standard volume weighting) measure exponentially favors inflation in the non-eternal case but exponentially disfavors it in the eternal case.  The two exponential behaviors are different: in the non-eternal case only the difference in the number of $e$-folds appears in the exponent while in the latter the exponent contains, in addition to $N_+ - N_-$, a large factor $H_0/H_s$ coming from the high energy of the false vacuum relative to the energy scale of inflation.

	\item The scale factor cutoff measure behaves the same as the proper time measure in the non-eternal case.  But in eternal inflation case, it does not care much about the duration of the  slow roll inflation.

	\item The causal diamond measure is like the scale factor cutoff in the eternal case: no reward for inflation.  But in non-eternal inflation it has an interesting behavior being sensitive to the number of $e$-folds only up to a critical number.  Beyond that the observer doesn't distinguish longer stages of inflation and hence the measure is insensitive to $N$.

	\item The prediction of the stationary measure coincides with the unambiguous counting of galaxies in non-eternal inflation.  This is shared by the previous two measures as well.  But the stationary measure gives the same result in both eternal and non-eternal case.  It is the only measure among the four we studied here that produces a result that continuously varies as one goes from eternal to non-eternal inflation.

\end{itemize}

It is quite interesting that all of these measures, being applied to non-eternal inflation, tell us that the probability to live in a given part of the universe is proportional to the exponential growth of volume during the slow-roll inflation, or at least during a certain part of inflationary expansion of the universe. Surprisingly, the first three of the measures in the Table \ref{tab:1} lost this universal property when applied to the eternal inflation scenario.

These results do not necessarily disfavor the first three measures since they have been invented for eternal inflation rather than for the non-eternal inflation. Nevertheless, dramatic discontinuity of predictions during the transition from non-eternal to eternal inflation is quite intriguing. With respect to the first two measures in the Table \ref{tab:1}, this discontinuity can be traced back to the use of the asymptotic stationary distributions at the stage when the stationarity is not reached for some of the processes, see a discussion of this issue in . Once this problem is taken care of \cite{Linde:2007nm,Linde:2008xf}, the transition from the non-eternal to eternal inflation becomes continuous. It would be interesting to see whether a similar modification can restore the continuity of predictions of other probability measures. 

\

\section*{Acknowledgment}
We are grateful to Daniel Harlow, Vitaly Vanchurin, Alex Vilenkin for very useful discussions.  This work was supported in part by NSF grant PHY-0244728. The work of A.L.\ was also supported by the FQXi grant RFP2-08-19. M.N.\ was supported in part by a Mellam Family Foundation Fellowship.

\end{document}